 \documentclass[final,5p,times,twocolumn,numbers]{elsarticle}

\usepackage{amssymb}
\usepackage{lipsum}
\usepackage{color}
\usepackage{ulem}
\biboptions{sort&compress}
\usepackage{amsmath}

\usepackage[utf8]{inputenc}
\usepackage[T1]{fontenc}

\makeatletter
\def\ps@pprintTitle{%
 \let\@oddhead\@empty
 \let\@evenhead\@empty
 \def\@oddfoot{}
 \let\@evenfoot\@oddfoot}
\makeatother

\begin{document}

\begin{frontmatter}



\title{Nuclear non-resonant photoexcitation assisted by electron recombination}


\author[1,2,3]{Nan Xue\fnref{equal}}
\author[3]{Zuoye Liu\fnref{equal}}
\fntext[equal]{These authors contributed equally to this work.}

\author[4]{Ziwen Li}
\author[5]{Adriana P\'alffy}
\author[6]{Jianmin Yuan}

\author[4]{Yuanbin Wu\corref{cor1}}
\ead{yuanbin@nankai.edu.cn}

\author[1,2]{Xiangjin Kong\corref{cor1}}
\ead{kongxiangjin@fudan.edu.cn}

\author[1,2,7]{Yu-Gang Ma\corref{cor1}}
\ead{mayugang@fudan.edu.cn}

\cortext[cor1]{Corresponding authors.}

\address[1]{Key Laboratory of Nuclear Physics and Ion-Beam Application (MOE), Institute of Modern Physics, Fudan University, Shanghai 200433, China}

\address[2]{Research Center for Theoretical Nuclear Physics, NSFC and Fudan University, Shanghai 200438, China}

\address[3]{School of Nuclear Science and Technology, Lanzhou University, Lanzhou 730000, China}

\address[4]{School of Physics, Nankai University, Tianjin 300071, China}

\address[5]{Institute of Theoretical Physics and Astrophysics, University of W\"urzburg, Am Hubland, W\"urzburg 97074, Germany}

\address[6]{Institute of Atomic and Molecular Physics, Jilin University, Changchun 130012, China}

\address[7]{School of Physics, East China Normal University, Shanghai 200062, China}

\begin{abstract}
We investigate theoretically a nuclear excitation mechanism involving absorption of non-resonant photons leveraged by the coupling to the atomic shell. The nuclear non-resonant photoexcitation is assisted by electron recombination which compensates the energy mismatch between photon and nuclear transition energies, reminiscent of parametric up-conversion in non-linear media. This third-order process proceeds via a virtual nuclear state-rather than virtual electronic states-distinguishing this mechanism from the electronic bridge. We investigate the process on the example of a so-far not observed $14.2$ keV hard x-ray transition in $^{193}\mathrm{Pt}$ driven by an x-ray free-electron laser. Although the calculated cross section is small, it can be compensated by the vast number of non-resonant photons from the x-ray laser pulse. By enabling nuclear excitation through non-resonant photons, this up-conversion-like mechanism suggests new directions for non-linear x-ray interactions mediated by nuclear transitions.

\end{abstract}
%\textcolor{red}{A related process is the electron bridge mechanism, distinguished by its coupling through electronic virtual states, whereas our mechanism proceeds via nuclear virtual states.}



\end{frontmatter}

%\tableofcontents



\section{Introduction}
\label{introduction}

Nuclei with narrow resonances, driven by advanced x-ray sources such as synchrotron radiation facilities and x-ray free-electron lasers (XFELs), serve as an interesting platform for quantum optics~\cite{NQO1,NQO3,rohlsberger2021quantum}. Due to their exceptionally narrow linewidths relative to the transition frequency~\cite{rohlsberger2004nuclear,NFS1}, nuclear transitions are well-suited for high-resolution measurements, enabling observations such as the collective Lamb shift~\cite{exp1} and the spontaneously generated coherence (SGC)~\cite{exp3}. Meanwhile, a variety of intriguing phenomena based on nuclear forward scattering (NFS) have been reported~\cite{rohlsberger2012electromagnetically,heeg2015tunable,heeg2015interferometric,haber2016,haber2017rabi,vagizov2014coherent,evers2017,heeg2021coherent,bocklage2021coherent,velten2024nuclear}, including the storage of nuclear excitation~\cite{shvyd1996storage} and superradiant emission from a nuclear ensemble~\cite{Chumakov2018}.
NFS experiments have also benefited from the development of phase-retrieval methods~\cite{yuan2025nuclear,negi2025energy,wolff2023unraveling}. 
%\sout{At the same time, nuclear transitions have been explored as potential timekeepers, with significant efforts directed toward developing a nuclear clock~\cite{Sc2023,PhysRevLett.132.182501,PhysRevLett.133.013201,Zhang2024}—an innovation that promises a next-generation framework for precision metrology and advances in fundamental physics~\cite{peik2021nuclear}.} 
At the same time, nuclear transitions have been explored as potential timekeepers, with significant efforts directed toward developing a nuclear clock ~\cite{Sc2023,PhysRevLett.132.182501,PhysRevLett.133.013201,Zhang2024,addTh1,addTh2,addTh3,Si2025,cpl_42_12_120302}—an innovation that promises a next-generation framework for precision metrology and advances in fundamental physics~\cite{peik2021nuclear}.

However, the same narrow linewidths that enable high precision also pose a practical challenge: only a tiny fraction of photons in an x-ray pulse are resonant with the transition energy. For example, in a recent experiment using the European XFEL to excite $^{45}$Sc, only about $10^{-15}$ of the photons in each pulse contributed to resonant excitation~\cite{Sc2023}. This mismatch greatly limits excitation efficiency, especially for transitions with unknown resonance energies, where wide-range energy scans are required. In a similar context, several approaches based on electronic bridge (EB) excitation mechanism have been proposed~\cite{EBexc1,EBexc2,EBexc3,EBexc4,EBexc5,Li2023}, in which the electronic shell couples to the nucleus and drives the nuclear excitation in conjunction with a photon. Extensions of the EB framework have also been explored, including—but not limited to—mechanisms involving continuum-state electrons~\cite{EBexc_new1,PhysRevC.102.024604}, laser-assisted schemes where external optical fields modulate the electronic coupling~\cite{EBexc_new2}, the two-photon EB mechanism~\cite{Cai2021}, and the hyperfine EB scheme~\cite{PhysRevLett.133.223001}. Similar electron-mediated schemes have also been proposed for the excitation of narrow-linewidth nuclear isomers, including laser-driven electron recollision~\cite{PhysRevLett.127.052501,PhysRevC.106.024606}, inelastic electron scattering~\cite{PhysRevC.106.044604,PhysRevC.106.064604}, and laser-driven cluster excitation~\cite{PhysRevLett.130.112501,PhysRevC.110.L051601}.

\begin{figure*}[t]
	\setlength{\abovecaptionskip}{-0.2cm} 
     \centerline{\includegraphics[width=1.00\linewidth]{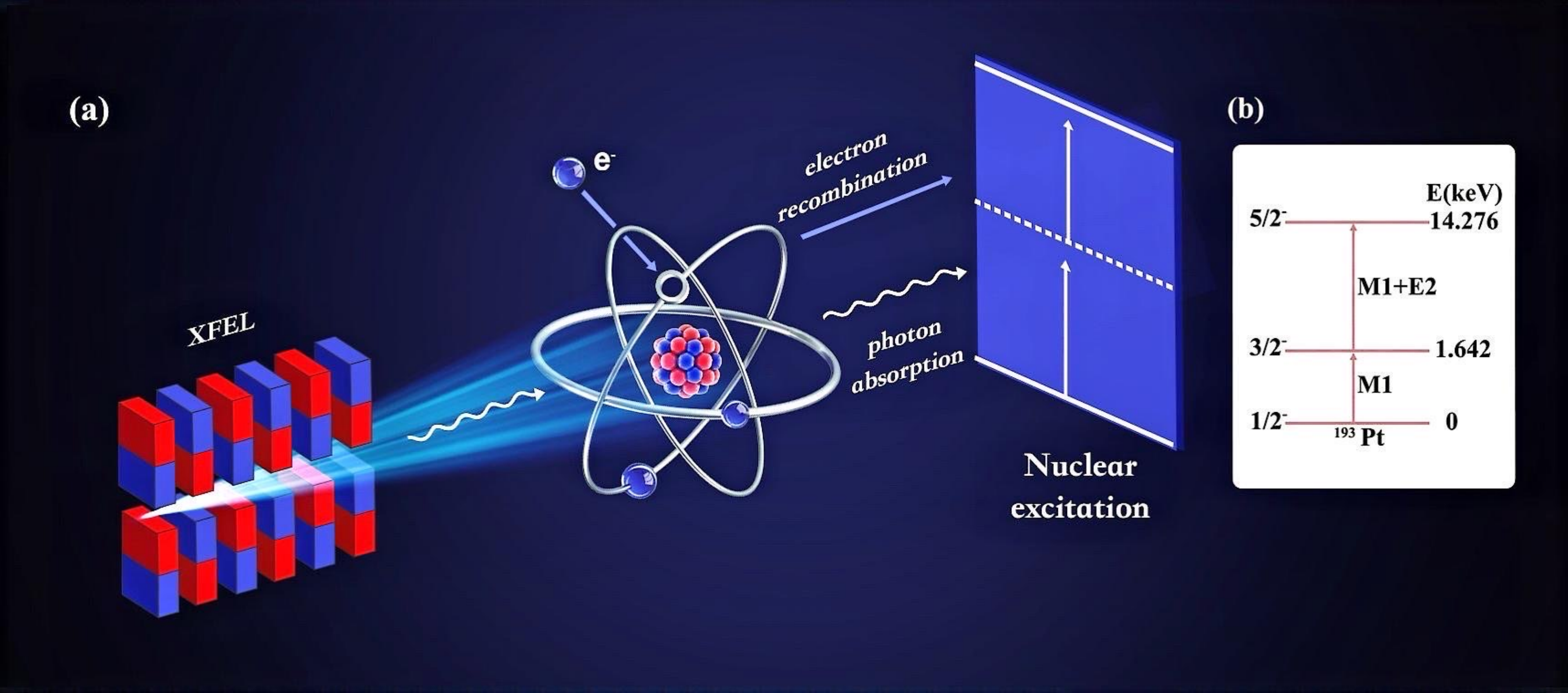}}
     \centering
	\begin{picture}(10,5)
     \end{picture}
	\caption{(a) Sketch of the process of photoexcitation assisted by electron recombination and (b) partial level scheme of $_{}^{193}\textrm{Pt}$ nucleus. The $_{}^{193}\textrm{Pt}$ level data is obtained from Ref.~\cite{ENSDF}.}
	\label{fig1} 
\end{figure*}

Here, we propose a nuclear excitation mechanism that allows non-resonant photons to contribute via a recombination-assisted process. In this mechanism, an incident x-ray photon and a free electron combine to excite the nucleus, effectively bridging the detuning between photon energy and the nuclear transition via a virtual nuclear state—rather than virtual electronic states—distinguishing our approach from the EB process. The process is third-order and has a lower cross section compared to direct photoexcitation. However, the abundance of non-resonant photons can compensate for this limitation, enabling potentially observable excitation rates. Conceptually, this mechanism resembles parametric up-conversion~\cite{upconversion1,upconversion2}: a recombining electron and a detuned photon jointly supply the energy needed to excite the nucleus, which acts as a non-linear medium. To explain the mechanism, we illustrate it on the example of a so-far unobserved nuclear transition in $^{193}\mathrm{Pt}$ and evaluate the expected excitation rates. For our numerical calculations, we consider XFEL photons in conjunction with a cold plasma which facilitates electron recombination. While the absolute rates are modest, the process represents a new conceptual route for nuclear excitation, complementing existing schemes.

\section{Theoretical Approach}
\label{Methods}

In our excitation mechanism, the nucleus is excited from the initial state $|I_{i}M_{i}\rangle$ to the final excited state $|I_{e}M_{e}\rangle$, by the absorption of a photon in conjunction with electron recombination into the ion. Here, $I_{i(e)}$ and $M_{i(e)}$ denote the total angular momentum and magnetic quantum number of the initial (excited) state, respectively.  This is a third-order process that proceeds via an intermediate nuclear virtual state $|I_{d}M_{d}\rangle$. At first, the nucleus undergoes a transition from the ground state to an intermediate virtual state via photon absorption. Subsequently, the nucleus is excited from the intermediate virtual state to the final excited state through electron recombination. Alternatively, the sequence of photonic and electronic processes can occur in inverse order. The total energy conversion caused by photon absorption and electron recombination must match the energy required for the nuclear transition. Fig.~\ref{fig1} presents a schematic diagram of this mechanism, showing that the nucleus is excited through x-ray photon absorption accompanied by electron recombination.

In this work, we consider an interesting case of $_{}^{193}\textrm{Pt}$, where the direct decay from the second excited state to the ground state has not been observed. The relevant energy levels of $^{193}\textrm{Pt}$ are presented in Fig.~\ref{fig1}(b). We investigate the influence of the photon energy and electron temperature on the excitation efficiency. In the process, the initial state $|\psi _{ini}\rangle$ can be expressed as a direct product of the nuclear, electronic, and photonic state vectors: $|\psi _{ini}\rangle=|I_{i}M_{i}\rangle\otimes |\psi _{i}^{el}\rangle \otimes |\lambda k{L}'M\rangle$. It includes the nucleus in its initial state, the free electronic state $|\psi _{i}^{el}\rangle=|\vec{p}m_{s}\rangle$, characterized by its momentum $\vec{p}$ and spin projection quantum number $m_{s}$, and the photonic state $|\lambda k{L}'M\rangle$ with wave number $k$, total angular momentum ${L}'$ and projection $M$. $\lambda$ determines whether the photon-induced transition is an electric  $\left (\lambda=\mathcal{E}\right )$ or a magnetic transition $\left (\lambda=\mathcal{M}\right )$.  The final state $|\psi _{fin}\rangle$ includes the nucleus in its excited state, the bound electron state $|\psi _{f}^{el}\rangle=|n_{f}\kappa_{f}m_{f}\rangle$, labeled by the principal quantum number $n$, Dirac angular momentum $\kappa$, and magnetic quantum number $m$, and the vacuum state: $|\psi _{fin}\rangle=|I_{e}M_{e}\rangle\otimes |\psi _{f}^{el}\rangle \otimes |0\rangle$. The intermediate state $|\psi _{mid}\rangle$ can take two forms, $|\psi _{mid}\rangle=|I_{d}M_{d}\rangle\otimes |\psi _{i}^{el}\rangle \otimes |0\rangle$ or $|I_{d}M_{d}\rangle\otimes |\psi _{f}^{el}\rangle \otimes |\lambda k{L}'M\rangle$.

The differential cross section of the process of nuclear photoexcitation assisted by electron recombination is given by
\begin{equation}
\frac{\mathrm{d}\sigma \left (E_{k},E_{p} \right )}{\mathrm{d} \Omega _{p} } =\frac{2\pi }{F_{i} } \left | \left \langle \psi_{fin}  \right | \hat{T}  \left (E_{k},E_{p} \right )\left | \psi_{ini}   \right \rangle  \right |^{2}\rho _{f} ,
\label{dsigma}
\end{equation}
where $E_{k}$ and $E_{p}$ are the kinetic energy of the continuum electron and the photon energy, respectively. Furthermore, $\Omega _{p}$ is the solid angle of the incoming electrons, $F_{i}$ stands for the incoming electron flux, $\rho _{f}$ represents the density of the final states, and $\hat{T}  \left (E_{k},E_{p} \right )$ is the third-order transition operator for the process. The electron flux $F_{i}$ can be expressed as $F_{i} \rho _{i} =\frac{p^{2} }{\left (2\pi \right )^{3} }$, where $\rho _{i}$ is the initial density of continuum electron states and $p$ is the value of the free electron momentum. By expanding the continuum electron states in terms of partial waves~\cite{Dirac}, and including both the nucleus–photon and nucleus–electron interactions, we obtain the cross section for a specific recombination channel, denoted by  $\alpha _{c}$, which labels the bound electronic orbital into which the continuum electron is captured 
\begin{eqnarray}
\sigma^{\alpha _{c} } \left (E_{k},E_{p} \right ) &=& \frac{2\pi}{F_{i}}\frac{1}{2\left (2I_{i}+1\right )\left (2{L}'+1\right )} \nonumber \\
&\times& \sum_{M_{i} }^{} \sum_{M_{e}m_{f}  }^{} \sum_{M }^{}\sum_{\kappa m }^{}\frac{1}{4\pi}\left |\Lambda _{1}+\Lambda _{2}\right |^{2},
\label{sigma}
\end{eqnarray}
where
\begin{eqnarray}
\Lambda _{1} &=& \sum_{\left | \psi_{\mathrm{I_{d}}  }   \right \rangle   }^{}\frac{1}{2I_{d}+1 }\label{matrixl1} \\
& \times & \sum_{M_{d} }^{} \Big[\frac{\alpha \left (\mathcal{M}(\mathcal{E}) {L}',I_{i}, M_i, I_{d}, M_d\right )}{E_{p}-\left (E_{n}^{d} -E_{n}^{i}\right )+\frac{i}{2}\Gamma _{d}} \nonumber\\
& \times & \frac{\sqrt{\Gamma _{f}/2\pi  }\left \langle I_{e}M_{e},n_{f}\kappa _{f}m_{f}  \right |H_{ne}^{\mathcal{M}(\mathcal{E})} \left | I_{d}M_{d},\varepsilon \kappa m   \right \rangle  }{E_{k}-\left (E_{n}^{f}-E_{p}-E_{bind}^{\alpha _{c} }\right )+\frac{i}{2}\Gamma _{f} }\Big], \nonumber
\end{eqnarray}
\begin{eqnarray}
\Lambda _{2} &=& \sum_{\left | \psi_{\mathrm{I_{d}}  }   \right \rangle   }^{}\frac{1}{2I_{d}+1 } \label{matrixl2} \\
& \times& \sum_{M_{d} }^{} \Big[ \frac{\alpha \left (\mathcal{M}(\mathcal{E}) {L}',I_{d},M_d, I_{e}, M_e \right )}{ (E_{n}^{f}-E_{n}^{d} )-E_{p}+\frac{i}{2}\Gamma _{f} } \nonumber\\
&\times & \frac{\sqrt{\Gamma _{d}/2\pi  }\left \langle I_{d} M_{d},n_{f}\kappa _{f}m_{f}  \right | H_{ne}^{\mathcal{M}(\mathcal{E})}\left | I_{i} M_{i} ,\varepsilon \kappa m \right \rangle   }{E_{k}-\left (E_{n}^{f} -E_{p}-E_{bind}^{\alpha _{c} }\right )+\frac{i}{2}\Gamma _{d} } \Big] . \nonumber
\end{eqnarray}
Here, the term $\Lambda _{1}$ describes the process in which the photon–nucleus interaction first couples the initial state to an intermediate virtual state, followed by the electron–nucleus interaction that connects the intermediate state to the final state. In contrast, $\Lambda _{2}$ corresponds to the inverse interaction sequence. Both terms contain a summation over the intermediate virtual nuclear states $\left | \psi_{\mathrm{I_{d}}  }   \right \rangle $. $E_{n}^{i,(d,f)}$ is the nuclear state energy of the initial state $(i)$, intermediate state $(d)$, or excited state $(f)$, and $\Gamma _{d/f}$ is the energy width of the corresponding state. $E_{bind}^{\alpha _{c} }$ represents the binding energy of the recombination channel $\alpha _{c}$. $H_{ne}^{\mathcal{M}(\mathcal{E})}$ stands for the nucleus-electron interaction Hamiltonian. The superscript marks the interaction type $\mathcal{M}$ and $\mathcal{E}$ for magnetic and electric interactions, respectively. The transition matrix element $\alpha \left (\mathcal{M}(\mathcal{E}) {L}',I_{i}, M_i, I_{d}, M_d\right )$ for the photon-nucleus interaction, and the transition matrix element of $H_{ne}^{\mathcal{M}(\mathcal{E})}$ can be connected to the nuclear reduced transition probability~\cite{Gunstphd, threej, ring2004nuclear}. The details of the transition matrix elements can be found in Appendix~A. According to Eqs.~(\ref{sigma})-(\ref{matrixl2}), the cross section is maximized when the sum of the incident photon energy $E_p$ and the energy released upon electron recombination ($E_k + E_{\rm bind}^{\alpha_c}$) matches the nuclear transition energy ($E_n^f - E_n^i$), thereby fulfilling the energy conservation condition.

In our work, a plasma system is taken as a representative example. The setup involves an XFEL pulse irradiating the plasma; further details on the experimental implementation are provided in Section \textbf{Experimental Setup Discussion}. Free electrons in the plasma can recombine with available ions and transfer energy to the nucleus. The ion charge-state distributions are calculated using the radiative–collisional code FLYCHK~\cite{FLYCHK}. In addition, non-resonant photons from the XFEL simultaneously contribute to the nuclear excitation process.
%\textcolor{red}{Both} a plasma environment and an XFEL source are involved: the plasma provides free electrons and ions with required vacancies for electron recombination, and the XFEL source provides detuned photons for nuclear excitation. 
The reaction rate for the process can be obtained by integrating the product of the cross section and the flux distribution,
%\begin{equation}
%\lambda ^{\alpha _{c}}\left (T_{e},n_{e} \right )  =\iint \sigma^{\alpha _{c}} \left (E_{k},E_{p}\right )\phi \left (E_{k},T_{e},n_{e}\right )f(E_{p} )dE_{k}dE_{p},
%\end{equation}
\begin{equation}
\begin{split}
\quad\quad \lambda^{\alpha_c}(T_e, n_e) &= \iint \sigma^{\alpha_c}(E_k, E_p) \\
&\times \quad \phi(E_k, T_e, n_e) \, f(E_p) \, dE_k \, dE_p,
\end{split}
\label{ratelambda}
\end{equation}
where $f(E_{p})$ is the normalized photon energy distribution and $\phi \left (E_{k},T_{e},n_{e} \right )$ is the electron flux. In the plasma environment, the electron flux $\phi \left (E_{k},T_{e},n_{e} \right )$ is obtained by modeling the plasma with a relativistic Fermi-Dirac distribution of electron temperature $T_{e}$ and density $n_{e}$~\cite{Gunst2015}. Generally, the energy distribution of the XFEL pulses follows a Gaussian distribution with an energy width of several eVs. An energy difference of a few eVs hardly affects the electron distribution in a plasma under consideration. Thus, the electron flux can be treated as a constant $\phi^{E_{k}}\left (T_{e},n_{e} \right )$ in the integral. Therefore, we can obtain the resonance strength $S$ and the reaction rate as
\begin{eqnarray}
& & S ^{\alpha _{c}}  = \int \sigma^{\alpha _{c}}\left  (E_{k},E_{p} \right )f\left (E_{p}\right )dE_{k}dE_{p}, \\
& & \lambda ^{\alpha _{c}}\left (T_{e},n_{e}  \right ) = \phi^{E_{k}}  \left (T_{e},n_{e} \right )S^{\alpha _{c}}.
\end{eqnarray}

Since there is no experimental data for a direct de-excitation from the second excited state ($5/2^-$) of $_{}^{193}\textrm{Pt}$ to the ground state ($1/2^-$) available in the NNDC database~\cite{ENSDF}, populating the second excited state directly from the ground state is expected to be difficult. This limitation thereby highlights the potential value of exploring alternative excitation pathways, such as the one considered in this work.

\section{Numerical results and discussions}
\label{Results}

\begin{figure}[ht]
	\setlength{\abovecaptionskip}{-0.2cm} 
     \centerline{\includegraphics[width=0.93\linewidth]{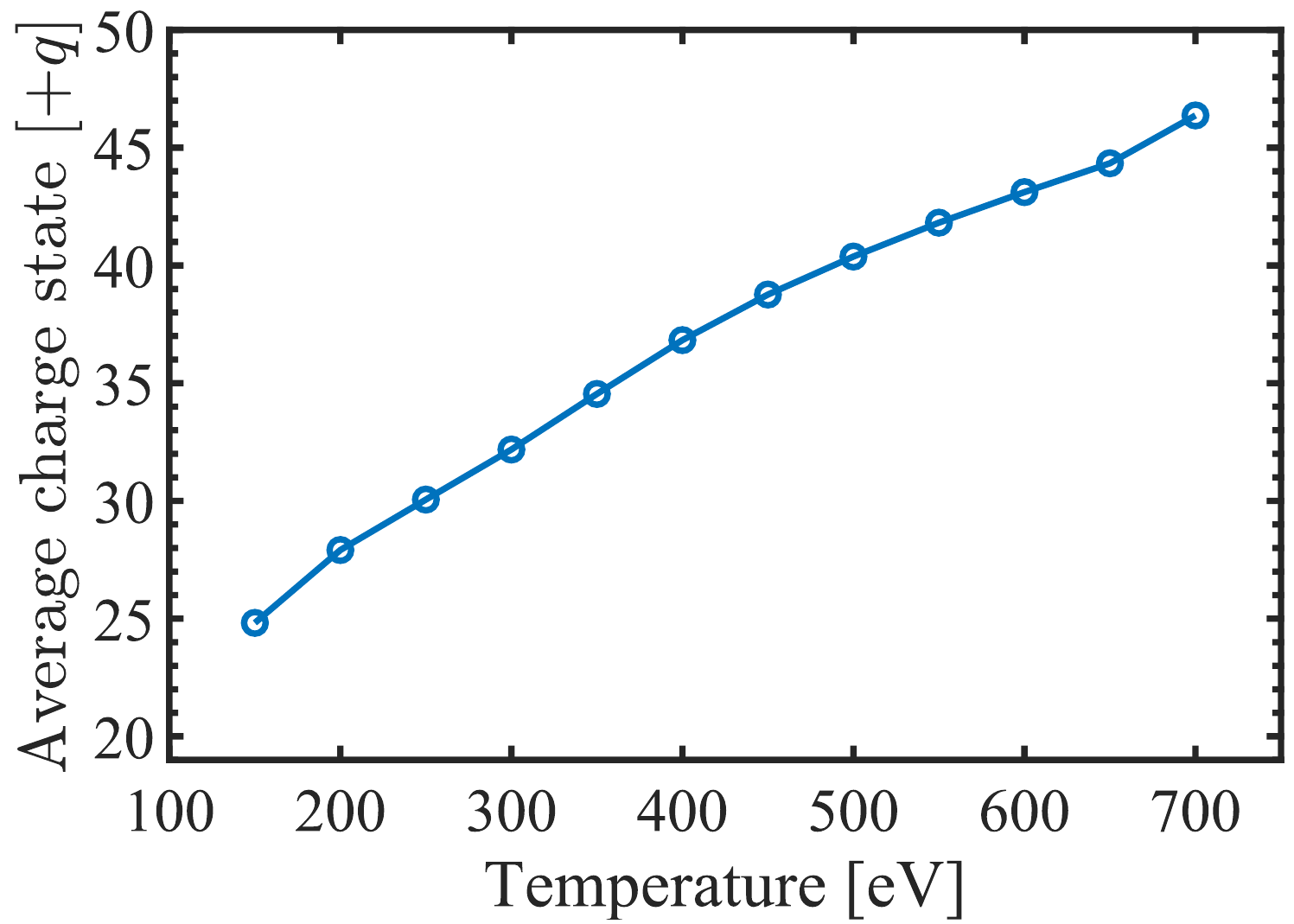}}
     \centering
	\begin{picture}(0,5)
     \end{picture}
	\caption{Average charge states $+q$ of ions at different electron temperatures, obtained with the FLYCHK code~\cite{FLYCHK}.}
	\label{CSD} 
\end{figure}

We assume here that the electron density of the plasma is $n_{e} =10^{24}$ cm$^{-3}$, which corresponds to a solid-state density. 
%The ion charge state distributions of the plasma at different electron temperatures are obtained by the radiative-collisional code FLYCHK~\cite{FLYCHK}. 
The initial electronic configuration before electron recombination is determined on the basis of the calculated average ion charge state presented in Fig.~\ref{CSD}. Our calculations consider the initial electronic state in the ground-state configuration and open recombination channels for the $s$, $p$, and $d$ subshells with the principal quantum number $n\le 6$ corresponding to the electron temperature. The total reaction rates and resonance strengths are obtained by the sum of all considered recombination channels. The electronic bound wave functions and binding energies are calculated using the relativistic multi-configurational Dirac-Fock method implemented in the code GRASP2K~\cite{Grasp2K}. The x-ray source intensity is assumed to be $10^{18}$ W/cm$^{2}$ with a pulse duration of $100$~fs. The required nuclear data including the excitation energy and reduced transition probabilities are obtained from Ref.~\cite{ENSDF}, and here we only consider the partial level scheme presented in Fig.~\ref{fig1}(b). The transition type between the $1/2^-$ and $3/2^-$ states is M1, whereas the transition between the $3/2^-$ and $5/2^-$ states is M1 $+$ E2, with the contribution of the E2 transition being very small (the mixing ratio $\delta$ is $0.015$)~\cite{ENSDF}. Furthermore, the next level above the second excited state $5/2^-$ lies as high as $114$ keV~\cite{ENSDF} and is energetically far away. Therefore, we restrict the sum over intermediate states to the state $3/2^-$ and consider only the M1 channel of the 12.634 keV transition. The reduced transition probabilities used in the calculations, $B(M1,5/2^{-}\to 3/2^{-})$ and $B(M1,3/2^{-}\to 1/2^{-})$, are taken as 0.0303~$W.u.$ and 0.165~$W.u.$, respectively~\cite{ENSDF}.

\begin{figure}[t]
	\setlength{\abovecaptionskip}{-0.2cm} 
     \centerline{\includegraphics[width=1.02\linewidth]{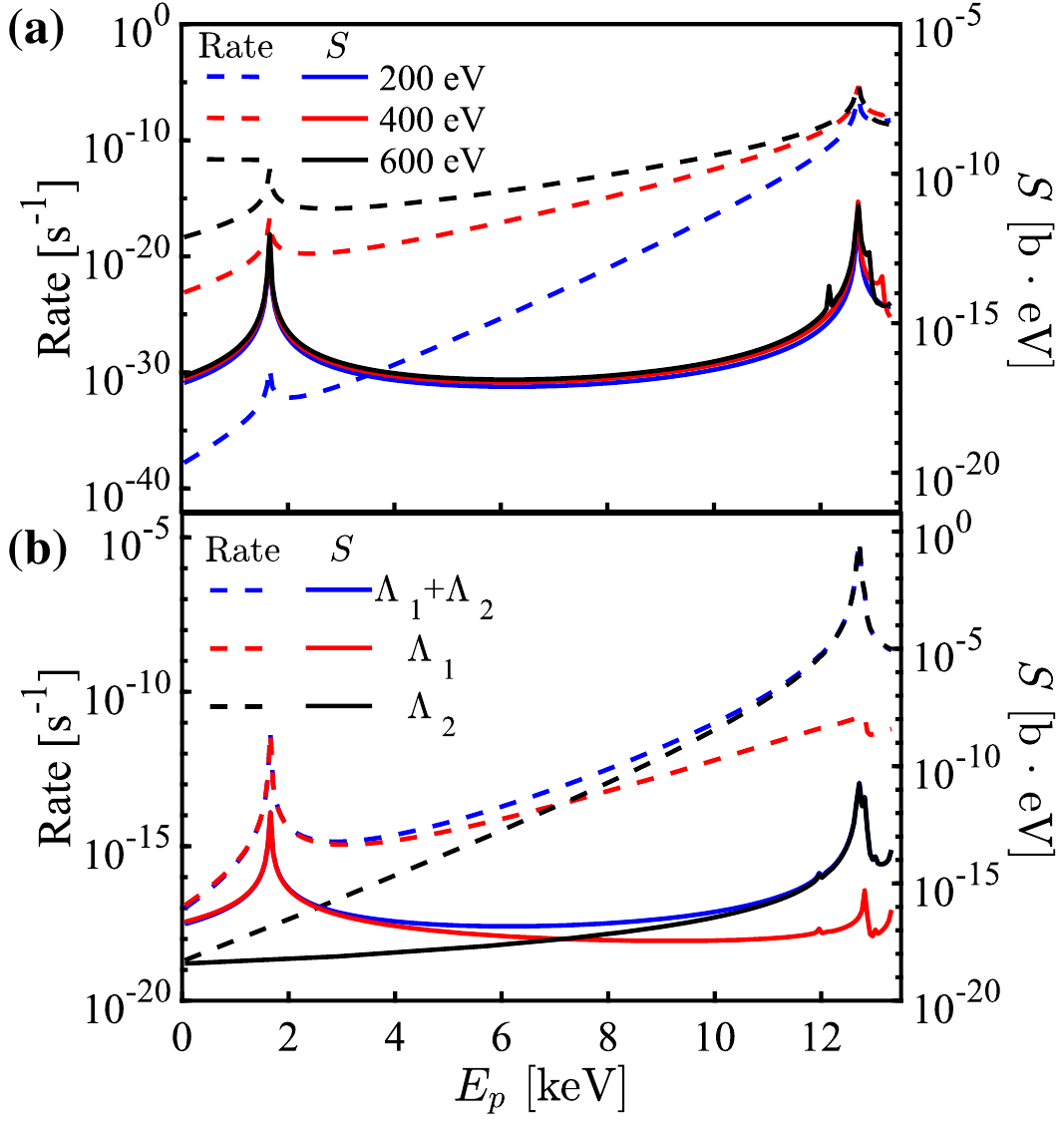}}
     \centering
	\begin{picture}(0,5)
     \end{picture}
	\caption{(a) Reaction rates and resonance strengths $S$ as functions of photon energy $E_p$ for electron temperatures $T _{e}$ = $200$ eV, $400$ eV, and $600$ eV. (b) Reaction rates and resonance strengths $S$ as functions of photon energy $E_p$ at $T_e = 700$ eV, shown for three contributions: including both $\Lambda_{1}$ and $\Lambda_{2}$, including only $\Lambda_{1}$, and including only $\Lambda_{2}$.}
	\label{rate_S}  
\end{figure}

The reaction rates and resonance strengths at selected electron temperatures are presented in Fig.~\ref{rate_S}(a). In our calculations, the photon energy $E_{p}$ is scanned from $0.05$~keV to $13.25$~keV with a step size of $0.05$~keV, which means the detuning at points near the resonance is still much larger than the natural linewidth of the nuclear energy levels. As the photon energy approaches the resonance energies of the two transitions ($1.642$~keV and $12.634$~keV), peaks appear in the reaction rates and resonance strengths. Different electron temperatures cause slight changes in the resonance strength by affecting the electronic configuration of the ions before electron recombination. As $E_{p}$ increases, the energy provided by electron recombination necessary to drive the nuclear transition gradually decreases. When this energy falls below the binding energy of the relevant orbital, the corresponding capture channel closes, resulting in fluctuations of the resonance strength at high $E_{p}$. The electron temperature primarily affects the flux distribution of high-energy electrons. Therefore, when $E_{p}$ is relatively high, the electron temperature does not significantly impact the reaction rate. This characteristic highlights the chance to reduce the control of plasma conditions, particularly in contrast to the typical electron recombination–induced nuclear excitation mechanism, namely nuclear excitation by electron capture (NEEC)~\cite{NEECini,Yang2025}. In plasma environments, conventional NEEC typically requires high-temperature plasmas to open inner-shell capture channels and provide a sufficient flux of resonant electrons~\cite{Gunst2015,plasma6,plasma8}, which substantially increases experimental complexity and background noise. Our scheme circumvents this constraint by introducing an additional photon to reduce the energy requirement for recombining electrons, enabling efficient nuclear excitation even in low-temperature plasmas while strongly suppressing background contributions from collisions and atomic processes.

The broadband electron energy distribution in the plasma, which may include a small fraction of high-energy electrons, may allow the direct driving of the nuclear excitation from the ground state to the $5/2^-$ state via the NEEC process, thereby necessitating a direct comparison with the conventional direct NEEC pathway. We assume a reduced transition probability $B_{3\to 1} (E2)=1$ $W.u.$ for the direct decay from the $5/2^-$ state to the ground state, as no data for this de-excitation is available in the NNDC database~\cite{ENSDF}. We consider the same $n_{e}$ and electron recombination channels, and calculate the NEEC reaction rate corresponding to the direct excitation from the ground state to the second excited state according to the method in Refs.~\cite{Palffy2006,Gunst2015}. We obtain $\lambda _{\mathrm{NEEC}} (200~{\textrm{eV}})=4.43\times 10^{-26}$ s$^{-1}$, $\lambda _{\mathrm{NEEC}} (400~{\textrm{eV}})=2.81\times 10^{-11}$ s$^{-1}$, and $\lambda _{\mathrm{NEEC}} (600~{\textrm{eV}})=7.58\times 10^{-6}$ s$^{-1}$. The NEEC rate varies significantly with the electron temperature. Low electron temperatures cannot provide a sufficient number of high-energy electrons, resulting in a small reaction rate. As a comparison, in the excitation mechanism proposed in the present work, the energy contributed from the absorbed photon reduces the electron energy required for the reaction, resulting in a much higher rate at low temperatures as compared to the direct NEEC process. 

As shown in Eqs.~(\ref{matrixl1})-(\ref{matrixl2}), there are two contributing parts $\Lambda _{1}$ and $\Lambda _{2}$ to the cross section Eq.~(\ref{sigma}), corresponding to the two sequences of transition pathways for a given nuclear intermediate virtual state. The modulus square of their sum results in the appearance of interference terms. To illustrate the impact of the interference, we calculate three cases: with only $\Lambda _{1}$, with only $\Lambda _{2}$, and with both contributions included. The results are presented in Fig.~\ref{rate_S}(b). To more clearly demonstrate the effect of the cross terms in $\left |\Lambda _{1}+\Lambda _{2}\right |^{2}$ in Eq.~(\ref{sigma}), we consider here the case with a relatively high electron temperature of $700$~eV. At this temperature, the electron flux distribution is relatively flat, resulting in a narrower range of reaction rates. Near the first resonance transition energy of $1.642$ keV, $\Lambda _{1}$ makes the primary contribution, while near the second resonance transition energy of $12.634$ keV, $\Lambda _{2}$ is the dominant component as shown in Fig.~\ref{rate_S}(b). The term determining whether the effect of interference is constructive or destructive—corresponding to the Fano interference—is given by 
\begin{equation}
\mathbb{E}_{int} = \left[\frac{1 }{(E_{n}^{f}-E_{n}^{d})-E_{p}+\frac{i}{2}\Gamma _{f} } \right]^{*} \left[ \frac{1 }{E_{p}-(E_{n}^{d}-E_{n}^{i})+\frac{i}{2}\Gamma _{d} } \right].
\label{Eint}
\end{equation}
%\begin{equation}
%\begin{split}
%\quad\quad\mathbb{E}_{\mathrm{int}} =& \left[ \frac{1}{(E_n^f - E_n^d) - E_p + %\frac{i}{2} \Gamma_f} \right]^* \\
%& \times \left[ \frac{1}{E_p - (E_n^d - E_n^i) + \frac{i}{2} \Gamma_d} \right].
%\end{split}
%\label{Eint}
%\end{equation}
If $1.642$ keV$\le E_{p} \le $ $12.634$ keV, then $\mathbb{E}_{int}$ is positive, and the cross term in $\left |\Lambda _{1}+\Lambda _{2}\right |^{2}$ is positive, leading to an increase of the resonance strength and reaction rate. This interference effect becomes more pronounced when $E_{p}$ is far from resonance. When the photon energy is around $7$~keV, the reaction rate considering all matrix elements is three to four times higher than that considering only $\Lambda _{1}$ or only $\Lambda _{2}$, as shown in Fig.~\ref{rate_S}(b). In the region near the resonance energy, one of the matrix elements dominates while the contribution of the other is negligible, so the impact of the cross term resulting from their combination is minimal.

\begin{figure}[t]
	\setlength{\abovecaptionskip}{-0.2cm} 
     \centerline{\includegraphics[width=1.00\linewidth]{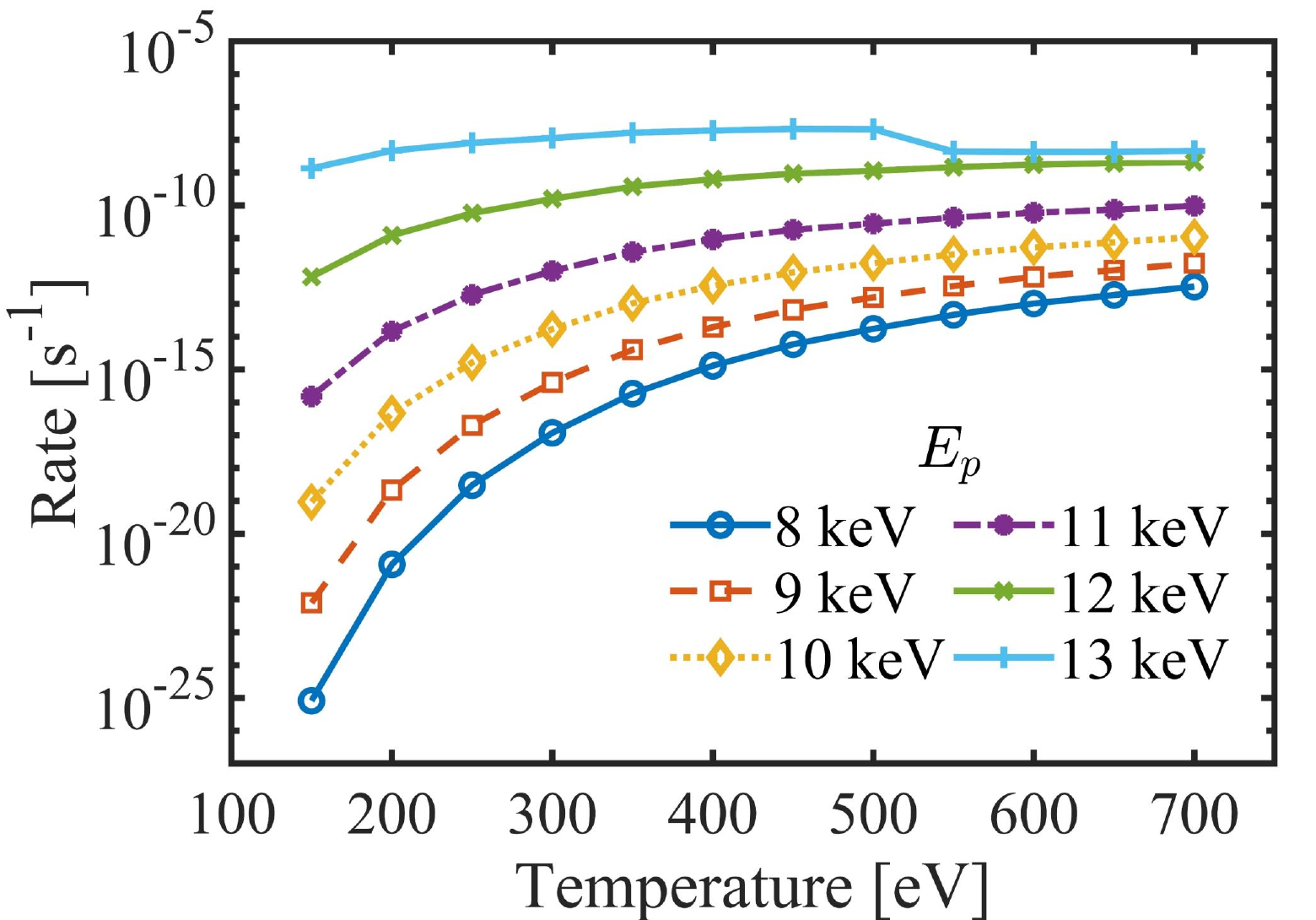}}
     \centering
	\begin{picture}(0,5)
     \end{picture}
	\caption{The reaction rate as a function of the electron temperature for selected photon energies.}
	\label{energy} 
\end{figure}

Now, we turn to analyse the dependence of the reaction rate on the electron temperature at given $E_{p}$, whose result is presented in Fig.~\ref{energy}. As the photon energy increases, the electron energy required for nuclear excitation decreases, allowing more electrons to participate in the reaction and thereby increasing the reaction rate. The electron temperature greatly affects the distribution of high-energy electrons while the impact on low-energy electrons is weaker. Therefore, scenarios requiring high-energy electrons will exhibit significant variations. When $E_{p}$ is sufficiently large, increasing the electron temperature results in little increase in the reaction rate, and it may even cause a decrease for $E_{p}=$ $13$ keV as shown in Fig.~\ref{energy}. This behaviour can be explained as follows. As the electron temperature increases, the charge state of the ions also increases, which in turn raises the binding energies of the recombination shells. The 5$p_{1/2}$ recombination channel, which significantly contributes to the overall reaction process, becomes closed as the binding energy exceeds the nuclear transition energy. The change in electron flux due to the increased electron temperature is insufficient to compensate for the loss of this channel, resulting in a decrease in the reaction rate when the temperature continues to rise beyond 500 eV for $E_{p}=$ $13$ keV.

\section{Experimental Setup Discussion}
\label{exp}

In the numerical calculations, we consider an experimental configuration involving both a plasma environment and an XFEL source. A natural platform is provided by beamlines that combine XFELs with intense optical lasers~\cite{xfel_opt1,xfel_opt2,xfel_opt3,xfel_opt4,HIBEF}, where the plasma is generated by the optical laser. Here, we consider a simplified implementation using only an XFEL source. In this scheme, the XFEL beam is split into two branches: one branch generates the plasma, while the other drives the excitation process after plasma formation~\cite{nature2012,PhysRevLett.109.065002,Lee:03}. The introduction of a photon in conjunction with electron recombination relaxes the plasma requirements, enabling our excitation scheme to operate in low-temperature plasmas directly generated by XFELs. This configuration eliminates the need for an additional intense optical laser and provides a cleaner environment with reduced background signals.  

Assuming a dense plasma with a temperature of $200$~eV, a total $^{193}$Pt ion number on the order of $10^{12}$, and a laser repetition rate of $100$ kHz, the event rate for nuclear excitation to the second excited state via the proposed process can reach approximately $10^{-3}\,\mathrm{s}^{-1}$ at a photon energy around 12.65 keV. Considering that the XFEL pulse duration of 100~fs determines the effective interaction time, which is much shorter than the typical plasma expansion timescale on the order of picoseconds~\cite{Gunst2015}, the effect of plasma expansion on the excitation efficiency is negligible. Regarding plasma generation, based on the experimental works of XFEL-generated plasmas in Refs.~\cite{nature2012,PhysRevLett.109.065002}, we estimate that achieving an electron temperature of around $200$ eV requires an intensity on the order of $10^{17}$~W/cm$^{2}$ for the X-ray pulses considered in this work, which should be within the reach of current XFEL facilities. In the actual experiment, since the incident XFEL photons are non-resonant with the nuclear transitions, they do not spectrally overlap with nor interfere with the signal photons of interest. Furthermore, the low temperature plasma created by the XFEL significantly suppresses background contributions from atomic and electronic processes, thereby providing a clean environment for detecting the nuclear excitation signal.

We note that, again assuming $B_{3\to 1} (E2)=1$ $W.u.$, direct photoexcitation from the ground state to the second excited state $5/2^-$ by an XFEL with the photon energy at the resonant energy of the transition between two levels could be more efficient compared with the excitation mechanism considered in the present work. In addition, the sequence of processes of NEEC from the ground state to the first excited state and then photoexcitation from the first excited state to the second excited state by an XFEL with the photon energy at the resonant energy of the transition between first excited state to the second excited state could also be more efficient compared with the excitation mechanism considered in the present work. However, these two excitation mechanisms mentioned above require that the photons in the x-ray pulse are resonant with the nuclear transition energy. Thus, the actual excitation efficiency could be reduced significantly by the necessary energy scans. 
%In addition, as one can apply an XFEL with the photon energy which is not at the resonant energy of the nuclear transition for the excitation mechanism considered in the present work, these two excitation mechanisms should not affect the observation of the proposed excitation mechanism. 
In addition, since the XFEL photon energy used in the present work is detuned from the nuclear transition resonance, these two excitation mechanisms should not affect the observation of the proposed scheme.

\section{Conclusions}
\label{conclusion}

In conclusion, we have introduced a new mechanism for nuclear excitation that combines non-resonant photon absorption with electron recombination. Conceptually reminiscent of up-conversion, the process involves the nucleus acting as a non-linear medium that absorbs joint energy from a photon and a recombining electron. This mechanism highlights an alternative excitation channel within the nuclear–electron–photon coupling framework. Demonstrated theoretically here for $^{193}\textrm{Pt}$, the concept may be extended to other systems and could motivate future research in non-linear physics involving nuclear transitions~\cite{Liu2025}, potentially contributing to advanced X-ray control techniques with implications for precision spectroscopy and related fields. Furthermore, the advent of higher-performance facilities, such as the Shanghai high repetition rate XFEL and extreme light facility (SHINE)~\cite{SHINE1,SHINE2,Yu2025}, will provide enhanced support for our proposed scheme.

\section*{Conflict of interest}
The authors declare that they have no conflict of interest.

\section*{Acknowledgements}
This work is supported by the National Key Research and Development Program of China under Contract No. 2024YFA1610900 and the National Natural Science Foundation of China (NSFC) under Grant No. 12447106, No. 12541501, No. 12475122, and No. 12547102. A.P. gratefully acknowledges the Heisenberg Program of the German Science Foundation (Deutsche Forschungsgemeinschaft, DFG) under Project No. 435041839.

\appendix

\section{Transition matrix elements.}
\label{AppendixA}

 The cross section of the photoexcitation process assisted by electron recombination in Eq.~(\ref{dsigma}) can be calculated by the perturbation expansion which leads to Eqs.~(\ref{sigma})-(\ref{matrixl2}). By applying the Wigner-Eckart theorem \cite{threej, ring2004nuclear}, the transition matrix elements in Eqs.~(\ref{matrixl1})-(\ref{matrixl2}) can be obtained. The transition matrix element $\alpha \left (\mathcal{M}(\mathcal{E}) {L}',I_{i}, M_i, I_{d}, M_d\right )$ for the photon-nucleus interaction is \cite{Gunstphd}
\begin{eqnarray}
& &\alpha \left (\mathcal{M}(\mathcal{E}) {L}',I_{i}, M_i, I_{d}, M_d\right )  \\
&=& \left (-1\right )^{I_{i}-M_{i}}C\left (I_{d}\; I_{i}\; {L}';M_{d} -\!\!M_{i}\;M \right ) \nonumber \\ & \times& E_{field}\sqrt{2\pi}\sqrt{\frac{{L}'+1}{{L}'}} \frac{k^{{L}'-1}}{\left (2{L}'+1\right )!!} \langle I_{d}||M_{L}(Q_{L})||I_{i} \rangle . \nonumber
\label{alpha}
\end{eqnarray}
Here $E_{field}$ is the electric field strength, and $C$ is the Clebsch–Gordan coefficient. The transition matrix element of $H_{ne}^{\mathcal{M}(\mathcal{E})}$ can be expressed as~\cite{Gunstphd}
\begin{eqnarray}
& & \langle I_{d}M_{d},n_{f}\kappa_{f}m_{f}|H_{ne}^{\mathcal{E}}|I_{i}M_{i},\varepsilon \kappa m\rangle \label{Hne1}\\
&=&\sum_{\mu =-L}^{L} \Big[ \left (-1\right )^{I_{d}+M_{i}+L+\mu +m+3j_{f}}\langle I_{d}||Q_{L}||I_{i} \rangle \nonumber\\ 
&\times& \sqrt{\frac{4\pi\left (2j_{f}+1\right ) }{\left (2L+1\right )^{3}}}  R_{L,\kappa_{f},\kappa}^{\mathcal{E}} C\left (I_{i}\; I_{d}\; L;-M_{i}\; M_{d}\;\mu \right )\nonumber\\
&\times& C\left (j\;j_{f}\;L;-m \;m_{f}-\mu \right )C\left (j_{f}\; L\; j;\frac{1}{2}\;0\;\frac{1}{2}\right ) \Big],  \nonumber
\end{eqnarray}
\begin{eqnarray}
& & \langle I_{d}M_{d},n_{f}\kappa_{f}m_{f}|H_{ne}^{\mathcal{M}}|I_{i}M_{i},\varepsilon \kappa m\rangle  \label{Hne2}\\
&=& \sum_{\mu =-L}^{L} \Big[ \left (-1\right )^{I_{i}-M_{i}+\mu +j-L-1/2}\langle I_{d}||M_{L}||I_{i} \rangle \nonumber\\ 
&\times& \sqrt{\frac{4\pi \left (2j+1\right )}{L^{2}\left (2L+1\right )^{2}}}\left (\kappa +\kappa _{f}\right ) C\left (j\;L\;j_{f};m -\mu\;m_{f}\right ) \nonumber\\
&\times& C\left (I_{d}\; I_{i}\; L;M_{d} -M_{i}\;\mu \right )\begin{pmatrix}
j_{f} & j & L \\
\frac{1}{2} & -\frac{1}{2} & 0
\end{pmatrix}R_{L,\kappa_{f},\kappa}^{\mathcal{M}} \Big]. \nonumber
\end{eqnarray}
Here $\binom{...}{...} $ is the Wigner 3j-symbol~\cite{threej}, $R_{L,\kappa_{f},\kappa}^{\mathcal{E}}$ and $R_{L,\kappa_{f},\kappa}^{\mathcal{M}}$ the radial integrals related to the coupling of electronic shells as defined in Ref.~\cite{Gunstphd}. The intermediate state $\left | \psi _{mid}   \right \rangle $ should include all cases that satisfy the selection rules. The electromagnetic element of the reduced transition matrix $\langle I_{d}||M_{L}(Q_{L})||I_{i} \rangle$ can be connected to the reduced transition probability by 
\begin{eqnarray}
&&\langle I_{d}||M_{L}(Q_{L})||I_{i} \rangle \\
&=&\sqrt{2I_{i}+1}\sqrt{B (\mathcal{M}(\mathcal{E}) {L},I_{i}\rightarrow I_{d})}e^{i\varphi \left ( I_{i}, I_{d} \right ) },\nonumber
\end{eqnarray}
where $|I_{i}\rangle$ and $|I_{d}\rangle$ are two different states, and $\varphi \left ( I_{i}, I_{d} \right )$ represents the phase introduced by the nuclear wavefunction. When considering only a single intermediate state, the phases of the two matrix elements are equal, allowing them to be eliminated through the modulus squared in the Eq.~(\ref{sigma}).

%\bibliographystyle{elsarticle-harv}
\bibliographystyle{elsarticle-num}
%\bibliography{example}
\bibliography{texref}



%% For example:

%% \bibitem[Aladro et al.(2015)]{Aladro15} Aladro, R., Martín, S., Riquelme, D., et al. 2015, \aas, 579, A101

%%\end{thebibliography}

\end{document}